\documentclass[aps,preprint,preprintnumbers,amsmath,amssymb,amsfonts,groupedaddress,superscriptaddress,showpacs,showkeys,floatfix]{revtex4-2}

\usepackage{graphicx}
\usepackage{subfigure}
\usepackage{bm}
\usepackage{amsmath}
\usepackage{color}
\usepackage{array}
\usepackage{CJK}
\usepackage{ulem}

\newcommand{\bmath}{\begin{mathletters}}
\newcommand{\emath}{\end{mathletters}}
\newcommand{\be}{\begin{eqnarray}}
\newcommand{\ee}{\end{eqnarray}}
\newcommand{\ba}{\begin{array}}
\newcommand{\ea}{\end{array}}

\newcommand{\pr}{\prime}

\newcommand{\rmS} {\mathrm{S}}

\begin{document}

\title{Experimental Determination of Multi-Qubit Ground State via a Cluster Mean-Field Algorithm}

 \author{Ze Zhan}
   \thanks{These authors have contributed equally to this work.}
 \affiliation{Zhejiang Province Key Laboratory of Quantum Technology and Device, Department of Physics, Zhejiang University, Hangzhou, 310027, China}
   \author{Chongxin Run}
     \thanks{These authors have contributed equally to this work.}
 \affiliation{Zhejiang Province Key Laboratory of Quantum Technology and Device, Department of Physics, Zhejiang University, Hangzhou, 310027, China}
  \author{Zhiwen Zong}
  \affiliation{Zhejiang Province Key Laboratory of Quantum Technology and Device, Department of Physics, Zhejiang University, Hangzhou, 310027, China}
\author{Liang Xiang}
 \affiliation{Zhejiang Province Key Laboratory of Quantum Technology and Device, Department of Physics, Zhejiang University, Hangzhou, 310027, China}
   \author{Ying Fei}
 \affiliation{Zhejiang Province Key Laboratory of Quantum Technology and Device, Department of Physics, Zhejiang University, Hangzhou, 310027, China}
   \author{Wenyan Jin}
 \affiliation{Zhejiang Province Key Laboratory of Quantum Technology and Device, Department of Physics, Zhejiang University, Hangzhou, 310027, China}
 \author{Zhilong Jia}
 \affiliation{Key Laboratory of Quantum Information, University of Science and Technology of China, Hefei, 230026, China}
 \author{Peng Duan}
 \affiliation{Key Laboratory of Quantum Information, University of Science and Technology of China, Hefei, 230026, China}
 \author{Jianlan Wu }
 \email{jianlanwu@zju.edu.cn}
 \affiliation{Zhejiang Province Key Laboratory of Quantum Technology and Device, Department of Physics, Zhejiang University, Hangzhou, 310027, China}
 \author{Yi Yin}
 \email{yiyin@zju.edu.cn}
 \affiliation{Zhejiang Province Key Laboratory of Quantum Technology and Device, Department of Physics, Zhejiang University, Hangzhou, 310027, China}
 \author{Guoping Guo}
 \email{gpguo@ustc.edu.cn}
 \affiliation{Key Laboratory of Quantum Information, University of Science and Technology of China, Hefei, 230026, China}
 \affiliation{Origin Quantum Computing, Hefei, 230026, China}

\begin{abstract}

A quantum eigensolver is designed under a multi-layer cluster mean-field (CMF) algorithm
by partitioning a quantum system into spatially-separated clusters. For each cluster,
a reduced Hamiltonian is obtained after a partial average over its environment cluster.
The products of eigenstates from different clusters construct a compressed Hilbert space,
in which an effective Hamiltonian is diagonalized to determine certain eigenstates of
the whole Hamiltonian. The CMF method is numerically verified in multi-spin chains
and experimentally studied in a fully-connected three-spin network,
both yielding an excellent prediction of their ground states.

\end{abstract}

\maketitle

{\it Introduction.} --- At the dawn of a quantum computing era, applications on quantum simulation and beyond have
attracted much attention of the whole quantum community. For example, mixed quantum-classical algorithms have
been proposed in the goal of solving unaffordable quantum chemistry problems with  quantum             
computers~\cite{PeruzzoNC14,MalleyPRX16,KandalaNat17,GoogleSci20,CollessPRX18,FarhiSci01,BarendsNat16,XChenPRL2010,chenxiPRAPP21,ZhanarXiv}.
A variational quantum eigensolver (VQE) was successfully implemented in the determination
of electronic states for a hydrogen molecule and multi-atom hydrogen chains~\cite{PeruzzoNC14,MalleyPRX16,KandalaNat17,CollessPRX18,GoogleSci20}.
The adiabaticity and shortcut-to-adiabaticity (STA) in analog and digitized designs~\cite{FarhiSci01,BarendsNat16,XChenPRL2010,chenxiPRAPP21}
can also be used in the quantum eigensolver,
where an eigenstate of the target Hamiltonian is obtained by dragging an eigenstate of an initial Hamiltonian through an adiabatic or STA trajectory.  
Recently, we proposed  a `leap-frog' algorithm via the digitized STA and adiabaticity~\cite{ZhanarXiv}.
Through a segmented trajectory of travelling intermediate states, our leap-frog method allows an
efficient and relibale  quantum eigensolver,  
as illustrated by our experimental study in H$_2$ and numerical calculation of hydrogen chains.

In the architecture of quantum computing, the eigenstructure of a $2^N$-dimensional ($2^N$-D) Hilbert space can be determined
in an $N$-qubit quantum device. However, the number of quantum gates in a digital quantum algorithm
quickly increases with the number of qubits~\cite{BarendsNC15,LloydSci96}.
In addition, a multi-qubit quantum gate is realized through a combination of single- and two-qubit gates~\cite{LloydPRL95}
but the number of the combining gates increases with the gate size.
The cost of quantum computing increases in company with the decrease of the fidelity
so that a practical quantum eigensolver is still limited by the system size.

In the fields of physics and chemistry, cluster-based methods have been applied on various 
problems~\cite{KadanoffPhy66,WilsonRMP75,YamamotoRRB09,HoyosPRB15,DrellPRD76,WhitePRL92,SchollwockRMP05}.
For example,  the concept of block spins was proposed to understand critical phenomena of the Ising model~\cite{KadanoffPhy66}.
In the renormalization group (RG) theory, the critical exponents are extracted from
the scale invariance around a fixed point~\cite{WilsonRMP75}. The clustering methods are also utilized in the quantum
chemistry computation~\cite{YamamotoRRB09,HoyosPRB15,DrellPRD76,WhitePRL92,SchollwockRMP05}.
In the block renormalization group (BRG) method, the total Hamiltonian is reconstructed in a compressed Hilbert space built by a few low-energy block states~\cite{DrellPRD76}.
In the density matrix renormalization group (DMRG), the compression of the Hilbert space is realized by the diagonalization
of reduced density matrices~\cite{WhitePRL92,SchollwockRMP05}.


In this paper, we will apply a multi-layer cluster mean-field (CMF) theory~\cite{HoyosPRB15} to build a new quantum eigensolver,
from which the eigenstructure of a large-scale system can be reliably and efficiently determined in a much smaller-scale quantum
device. The product states combined from the eigenstates of reduced cluster Hamiltonians define a compressed Hilbert space,
in which the effective Hamiltonian is diagonalized for the eigensolver. This CMF method is numerically verified in $N$-spin chains
and experimentally implemented in a fully-connected three-spin system, both
yielding high fidelities for the extracted ground states.


{\it Theory.} --- In a general multi-electron system, the second quantized Hamiltonian can be transformed into a multi-spin form,
\be
H = g^{(0)}+\sum_{i=1}^N\sum_{a=1}^3 g^{(1)}_{i;a} \sigma_{i}^{a}+\sum_{i,j=1}^N\sum_{a,b=1}^3 g^{(2)}_{ij;ab} \sigma_{i}^{a}\sigma_{j}^{b}
+\sum_{i,j,k=1}^N\sum_{a,b,c=1}^3 g^{(3)}_{ijk;abc} \sigma_{i}^{a}\sigma_{j}^{b}\sigma_{k}^{c}+\cdots,
\label{eq_01}
\ee
through a fermion-to-spin mapping method such as the Bravyi-Kitaev transformation~\cite{SeeleyJCP12}.
Here $\{\sigma_{i}^{a}\!=\!X_i, Y_i, Z_i\}$ is the set of the Pauli matrices acting on spin $i$
and the coefficients
$\{g^{(0)}, g^{(1)}, g^{(2)}, \cdots\}$ describe the strengths of (multi)-spin interactions.
To keep its generality, Eq.~(\ref{eq_01}) is allowed to include an arbitrary $N^\pr(\le N)$-spin interaction.

\begin{figure}[tp]
\centering
 \includegraphics[width=0.7\columnwidth]{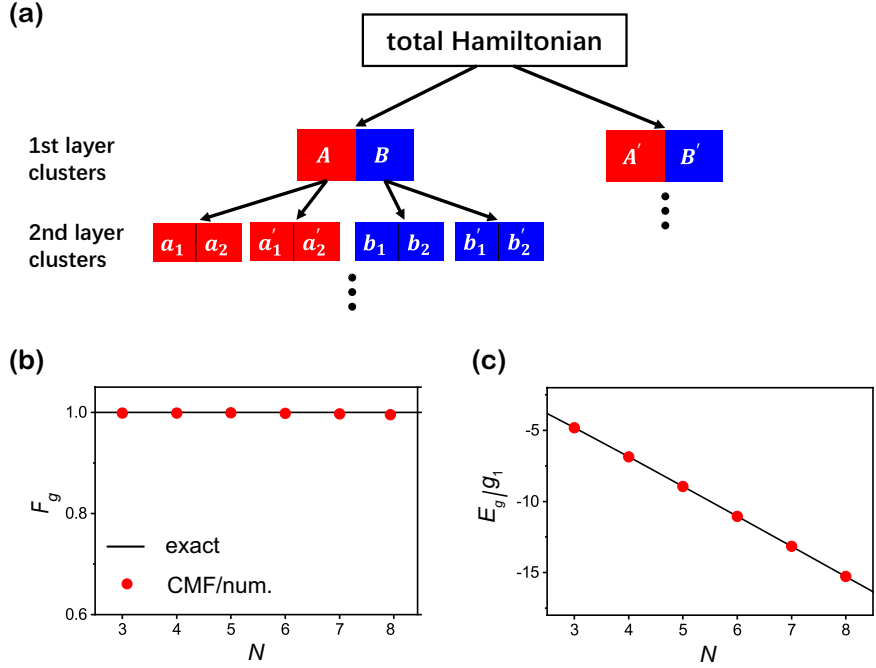}
\caption{(a) A schematic diagram of a multi-layer CMF algorithm.
(b-c) The numerical calculation of this CMF method for the $N$-spin systems [the Hamiltonian in Eq.~(\ref{eq_02})]:
(b) the fidelity $\mathcal F^\mathrm{theo}_g$ of the ground state  and (c) its eigenenergy $E_g$.
The red circles denote the numerical results of the CMF while the solid lines denote the exact results.
}
\label{fig_01}
\end{figure}

To extract the exact eigenstates and eigenenergies ($|\Psi_n\rangle$ and $E_n$) of the Hamiltonian in Eq.~(\ref{eq_01}), we require a diagonalization tool in
a $2^N$-D Hilbert space. Instead, a CMF method can realize an approximate but reliable
eigensolver in a highly compressed space.
For simplicity, we assume that an $N$-spin network is divided into two clusters, each with
$N_A$ and $N_B (=\!N\!-\!N_A)$ spins.
For a given $N_A$, the total choices of cluster partitioning
are $M_\mathrm{max}=C^N_{N_A}$ but a practical number $M$  can be much smaller than $M_\mathrm{max}$.

Next we interpret our CMF method as follows [see Fig.~\ref{fig_01}(a)]. For a given $A$-$B$ partition, we inspect the two clusters separately.
For cluster $A$, the rest part of the spin network (cluster $B$) is viewed as its environment.
After a partial trace over a specific $B$-state $|\varphi^\alpha_{B}\rangle$,  
a reduced Hamiltonian $H^\alpha_{A}=\langle\varphi^\alpha_B|H|\varphi^\alpha_{B}\rangle$ is constructed
and its diagonalization leads to a set of eigenstates $\{|\varphi^{i_\alpha}_{A}\rangle\}$ and
eigenenergies $\{\varepsilon^{i_\alpha}_{A}\}$, i.e., $H^\alpha_{A}=\sum_{i}\varepsilon^{i_\alpha}_{A}|\varphi^{i_\alpha}_{A}\rangle\langle\varphi^{i_\alpha}_{A}|$.
The same approach can be applied vice versa. With respect to an $A$-state $|\varphi^\beta_{A}\rangle$,
the reduced $B$-Hamiltonian $H^\beta_{B}=\langle\varphi^\beta_{A}|H|\varphi^\beta_A\rangle$ is diagonalized into
$H^\beta_{B}=\sum_{j}\varepsilon^{j_\beta}_{B}|\varphi^{j_\beta}_{B}\rangle\langle\varphi^{j_\beta}_{B}|$.
The two sets of product states, $\{|\varphi^{i_\alpha}_{A}\rangle\otimes|\varphi^\alpha_{B}\rangle\}$
and $\{|\varphi^\beta_{A}\rangle\otimes|\varphi^{j_\beta}_{B}\rangle\}$, from all the necessary cluster partitions
are mixed together to form a basis set of $\{|\psi_{\gamma}\rangle\}$ for a compressed Hilbert space. To capture a mean-field spirit,
we expect that all the states are self-consistently determined, i.e., $\{|\varphi^{i_\alpha}_{A}\rangle\}=\{|\varphi^\beta_A\rangle\}$
and $\{|\varphi^{j_\beta}_{B}\rangle\}=\{|\varphi^\alpha_B\rangle\}$. Although a regular recursive iteration is divergent if more than one states are
considered, the number of relevant states  is in general unchanged.
In practice, we take a limited number of iteration steps. At the final step, irrelevant states are discarded
and the Schmidt orthogonalizations~\cite{BjorckBIT67} is used to extract an orthonormal basis set $\{|\psi^\mathrm{S}_{\gamma}\rangle\}$.
An effective Hamiltonian,
\be
H_\mathrm{eff} = \sum_{\gamma\gamma^\prime} H_{\gamma\gamma^\prime}|\psi^\rmS_{\gamma}\rangle\langle\psi^\rmS_{\gamma^\prime}|
\label{eq_01a}
\ee
with $H_{\gamma\gamma^\prime}=\langle\psi^\rmS_{\gamma}|H|\psi^\rmS_{\gamma^\prime}\rangle$, is thus defined.
The digonalization of $H_\mathrm{eff}$ provides a good estimation of certain eigenstates $|\Psi_n\rangle$ and eigenenergies $E_n$.
If the number of the product states associated with each cluster partition is $J$, the dimensionality of
the compressed space is $MJ$, which can be significantly smaller than $2^N$.
The partition can be subsequently applied to clusters $A$ and $B$, e.g., $A=a_1\oplus a_2=a^\pr_1\oplus a^\pr_2=\cdots$,
which eventually leads to a multi-layer CMF algorithm [see Fig.~\ref{fig_01}(a)].
Relatively speaking, our CMF method takes a top-down strategy by partitioning a large system into small clusters
while the DMRG takes a bottom-up strategy by extending the system size with the increment of boundary spins.

{\it Numerical study.} --- To demonstrate the applicability of this CMF method, we numerically calculate the ground state $|\Psi_g\rangle$ and
its eigenenergy $E_g$ of an $N$-spin chain whose Hamiltonian reads~\cite{GoogleSci20}
\be
H = \sum_{i=1}^N g_1 Z_i +\sum_{i=1}^{N-1} g_2 X_i X_{i+1}.
\label{eq_02}
\ee
In our numerical calculation, the chain length is set to be $3\le N \le 8$ while the two parameters are fixed at $g_2/g_1=2$.
For each $N$-spin chain, we only consider two choices of cluster partitioning, $\{A=\{s_1, s_2\}, B=\{s_3,\cdots s_N\}\}$
and $\{A^\pr=\{s_1,\cdots, s_{N-2}\}, B^\pr=\{s_{N-1}, s_N\}\}$, where $s_i$ denotes the $i$-th spin.
Taking the first cluster partition as an example, we show the numerical approach in detail. In the first stage,
an initial $B$-state, $|\varphi_{B}\rangle\propto\prod \limits_{n=3}^N(|+\rangle+|-\rangle)_n$, is used to obtain a reduced $A$-Hamiltonian,
\be
H_A = \bar{\varepsilon}_A+ g_{1}Z_1+g_1Z_2+g^A_1X_2+ g_2X_1 X_2
\label{eq_03}
\ee
with $\bar{\varepsilon}_A=\sum_{n=3}^Ng_1\langle \varphi_B|Z_n|\varphi_B\rangle$ and $g^A_1=g_2\langle\varphi_B|X_3|\varphi_B\rangle$.
Due to the final goal of calculating $|\Psi_g\rangle$, we only select two $A$-eigenstates, the ground and first excited states of $H_A$,
i.e., $\{|\varphi_{A}^{\beta}\rangle=|\varphi_{A}^{g}\rangle,|\varphi_{A}^{e}\rangle\}$.
In the second stage, two $B$-Hamiltonians,
\be
H_{B}^{\beta=g, e}=\bar{\varepsilon}_B+ g^B_1X_3+\sum_{n=3}^N g_1 Z_n+\sum_{n=3}^{N-1} g_2 X_n X_{n+1},
\label{eq_04}
\ee
are extracted with respect to these two $A$-eigenstates. The two parameters are given by
$\bar{\varepsilon}_B=g_1\langle \varphi^\beta_A|Z_1+Z_2|\varphi^\beta_A\rangle$ and $g^B_1=g_{2}\langle\varphi_{A}^{\beta}|X_2|\varphi_{A}^{\beta}\rangle$.
The diagonalization of $H^{\beta=g, e}_B$ leads to four (ground and first excited) $B$-eigenstates, $|\varphi_{B}^{j_\beta}\rangle$ with $j=\{g,e\}$.
In the third stage, we use these four $B$-states $|\varphi^\alpha_B\rangle$ ($\alpha=g_g,g_e,e_g,e_e$) as the environment states
and calculate eight $A$-eigenstates $|\varphi^{i_\alpha}_A\rangle$ ($i=g, e$). To avoid the
divergence of this recursion,  we stop at this stage and discard four crossing terms. The four remaining products are
$|\varphi_{A}^{i_\alpha}\rangle\otimes|\varphi_{B}^{\alpha}\rangle$ with $\{i=g,\alpha=g_g,g_e\}$ and $\{i=e,\alpha=e_g,e_e\}$.
After including the four relevant product states from the second cluster partition $\{A^\pr, B^\pr\}$, the
Schmidt orthogonalization is applied to construct an 8-D Hilbert space.
The ground state $|\Psi_g\rangle$ and its eigenenergy $E_g$ are then determined by the diagonalization of
$H_\mathrm{eff}$ in Eq.~(\ref{eq_01a}). In the case of $N(> 4)$-spin chains, the multi-layer approach is
utilized so that all the calculations are restricted in the $N^\pr(\le 4)$-spin Hamiltonians. For example,
the total 50 two-spin, 5 three-spin and 16 four-spin Hamiltonians are involved for the 8-spin chain.

The numerical results are presented in Figs.~\ref{fig_01}(b) and~\ref{fig_01}(c).
Here we introduce a fidelity function, $\mathcal F^\mathrm{theo}_g\!=\!|\langle\Psi^\mathrm{exact}_g|\Psi^\mathrm{theo}_g\rangle|^2$,
between the CMF result $|\Psi^\mathrm{theo}_g\rangle$ and the exact state $|\Psi^\mathrm{exact}_g\rangle$.
As shown in Fig.~\ref{fig_01}(b), the CMF predictions
are excellent, satisfying $\mathcal F^\mathrm{theo}_g(3\le N\le 8)>99.4\%$. As shown in Fig.~\ref{fig_01}(c),
the accuracy of the ground state energy $E_g$ is even higher ($>99.9\%$) and a good linear dependence
is observed between $E_g$ and $N$.
As a comparison, we test a DMRG-based quantum eigensolver and the final state fidelities are around $99.0\%$. 


{\it Experimental study.} --- Next we use a two-qubit device to extract $|\Psi_g\rangle$ and $E_g$ of a fully-connected three-spin system
as an experimental demonstration of the CMF algorithm.
Due to the restriction of our current setup, it is difficult for us to reliably explore larger systems
which will be left in the future.
Our quantum device is composed of two superconducting cross-shaped transmon qubits~\cite{HouckNature07,BarendsPRL13,XLPRAPP2020}.
The ground and excited states of each qubit are one-to-one mapped onto the spin up and down states, i.e.,
$|0\rangle\!\leftrightarrow\!|+\rangle$ and $|1\rangle\!\leftrightarrow\!|-\rangle$.
The operation points of the two qubits are $\omega_a/2\pi\!=\!5.46$ GHz and $\omega_b/2\pi\!=\!4.92$ GHz,
while their anharmonicities are $\Delta_a/2\pi\!\approx\!\Delta_b/2\pi\!=\!-250$ MHz.
The relaxation times are $T_{a;1}\!=\!16.1$ $\mu$s and $T_{b;1}\!=\!26.5$ $\mu$s,
and the pure dephasing times are $T_{a;\phi}\!=\!20$ $\mu$s and $T_{b;\phi}\!=\!45$ $\mu$s.
The readout fidelities of the ground and excited states are $\{F_{a;0}\!=\!99\%, F_{a;1}\!=\!93\%\}$
and $\{F_{b;0}\!=\!96\%, F_{b;1}\!=\!94\%\}$.

\begin{figure}[tp]
\centering
 \includegraphics[width=0.7\columnwidth]{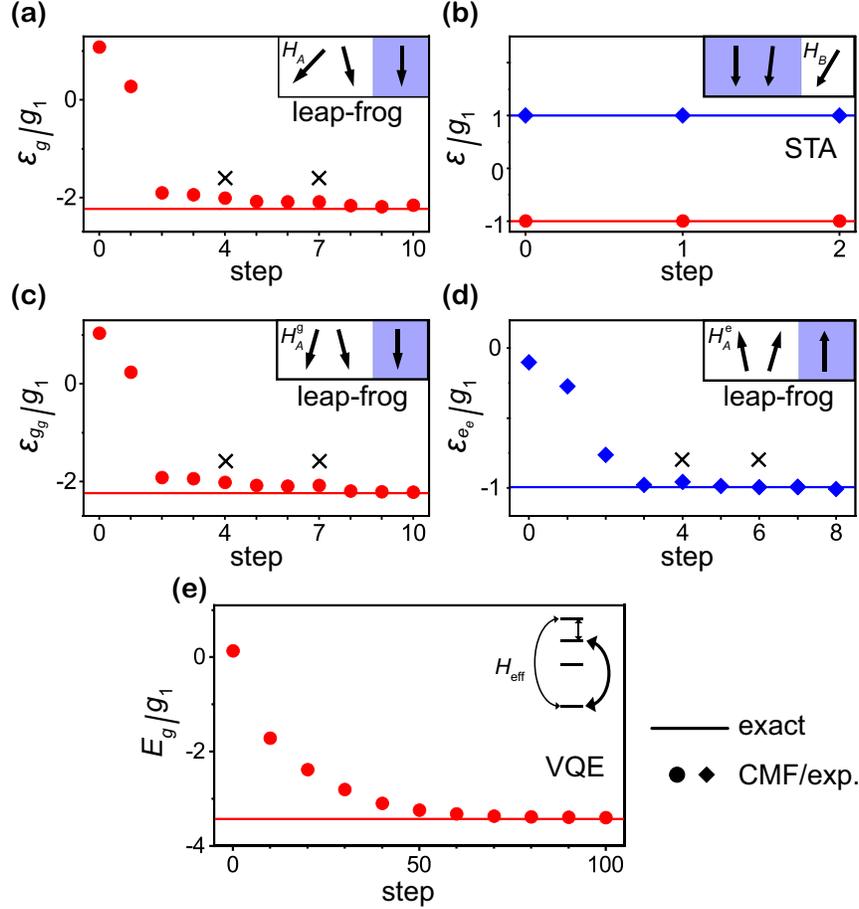}
\caption{A 4-stage CMF experiment to determine the ground state $|\Psi_g\rangle$ for the 3-spin Hamiltonian
in Eq.~(\ref{eq_06}) with $g_2/g_1=1.0$ and $g_3/g_1=0.1$.
(a) The 1st-stage eigenenergy evolution in a 3-segment leap-frog determination of $|\varphi^g_A\rangle$ for a reduced $A$-Hamiltonian $H_A$.
(b) The 2nd-stage eigenenergy evolutions in the digitized STA determination of $|\varphi^g_B\rangle$ and $|\varphi^e_B\rangle$
for the subsequent $B$-Hamiltonian $H_B$. (c-d) The 3rd-stage eigenenergy evolutions
in the leap-frog determinations of (c) $|\varphi^{g_g}_A\rangle$ and (d) $|\varphi^{e_e}_A\rangle$ for  $H^g_A$ and $H^e_A$, respectively.
The cluster partition is shown in the inset of each panel. In (a), (c) and (d), each cross labels an intermediate state in the leap-frog algorithm.
(e) The 4th-stage eigenenergy evolution in the VQE determination of $|\Psi_g\rangle$ for the 4-D effective Hamiltonian $H_\mathrm{eff}$.
The structure of $H_\mathrm{eff}$ is depicted in the inset. In each panel, the symbols denote the experimental results while the solid horizontal lines label their exact values.
}
\label{fig_02}
\end{figure}

The Hamiltonian of the three-spin system being studied is
\be
H =  g_1(Z_1+Z_2+Z_3)+g_2(X_1 X_2+X_2 X_3)+g_3 X_1 X_2 X_3,
\label{eq_06}
\ee
where the three-spin interaction $X_1 X_2 X_3$ increases the difficulty of the eigensolver.
In this paper,  two sets of experiments are performed to explore the influences of $g_2/g_1$ and $g_3/g_1$ separately.
In the first set, we fix $g_3/g_1=0.1$ and investigate $|\Psi_g\rangle$ and  $E_g$ upon the change of $g_2/g_1$.
For simplicity, we only consider three values, $g_2/g_1=0.1$, 1.0 and 2.0.
To visualize our experimental procedure, we take $g_2/g_1=1.0$ as an example and provide the stage-by-stage results in Fig.~\ref{fig_02}.
(i) We treat spins 1 and 2 as cluster $A$ and spin 3 as cluster $B$. With an initial guess of the $B$-state,
 $|\varphi_B\rangle=|1\rangle$, a reduced $A$-Hamiltonian is obtained as
$H_A = H^0_A+g^A_1X_2 + g^A_2X_1 X_2$ with $H^0_A=\bar{\varepsilon}_A+ g_{1}Z_1+g_1Z_2$.
Here the $B$-averaged parameters are $\bar{\varepsilon}_A=g_1\langle \varphi_B|Z_3|\varphi_B\rangle$,
$g^A_1=g_2\langle\varphi_B|X_3|\varphi_B\rangle$ and $g^A_2=g_{2}+g_{3}\langle\varphi_B|X_3|\varphi_B\rangle$.
The ground state of $H_A$ is experimentally determined by a leap-frog algorithm via the digitized
STA and adiabaticity~\cite{supp}.
With two varying parameters $\lambda_1$ and $\lambda_2$, the $A$-Hamiltonian is extended to be
\be
H_A(\lambda_1, \lambda_2) = H^0_A+\lambda_1g^A_1X_2 +\lambda_2 g^A_2X_1 X_2.
\label{eq_07}
\ee
As shown in Fig.~\ref{fig_02}(a), we begin with an initial Hamiltonian $H^0_A=H_A(\lambda_1=0, \lambda_2=0)$
and prepare its ground state $|\varphi^g_A(H^0_A)\rangle=|11\rangle$. A 4-step digitized
STA is applied to drag this state to the ground state $|\varphi^g_A(H^1_A)\rangle$
of an intermediate Hamiltonian $H^1_A=H_A(\lambda_1=0, \lambda_2=0.1)$.
Subsequently, two digitized adiabatic processes
realize an evolution of $|\varphi^g_A(H^1_A)\rangle\rightarrow|\varphi^g_A(H_A(\lambda_1=0, \lambda_2=0.5))\rangle \rightarrow |\varphi^g_A(H_A)\rangle$.
The theoretical prediction of the final state fidelity is $\mathcal F^\mathrm{theo}_g=99.9\%$ while the experimental determination is at $\mathcal F^\mathrm{exp}_g=98.6\%$.
(ii) In the second stage [see Fig.~\ref{fig_02}(b)], we input the previous $A$-state $|\varphi^g_A\rangle$
and calculate the   $B$-Hamiltonian,  $H_{B}=H^0_B+g^B_1 X_3$
with $H^0_B=\bar{\varepsilon}_B+g_{1}Z_3$. Here $\bar{\varepsilon}_B$ and $g^B_1$ are two $A$-averaged parameters.
The ground and excited states, $|\varphi^g_{B}(H_B)\rangle$ and $|\varphi^e_B(H_B)\rangle$, are
experimentally determined via the digitized STA from the two initial states $|\varphi^g_B(H^0_B)\rangle=|1\rangle$
and $|\varphi^e_B(H^0_B)\rangle=|0\rangle$. The experimental fidelities of these two $B$-eigenstates are $\mathcal F^\mathrm{exp}_g\approx 99\%$.
(iii) In the third stage, the two $B$-states $|\varphi^g_B\rangle$ and $|\varphi^e_B\rangle$
are used to obtain two $A$-Hamiltonians, $H^{\alpha=g, e}_A=\langle\varphi^\alpha_B|H|\varphi^\alpha_B\rangle$,
which are extended to the same form $H_A(\lambda_1, \lambda_2)$ as in Eq.~(\ref{eq_07})
but the $B$-averaged parameters are updated.
As shown in Figs.~\ref{fig_02}(c) and \ref{fig_02}(d), the leap-frog algorithm is also applied to experimentally determine
the ground state $|\varphi^{g_g}_A\rangle$ of $H^g_A$ and the first excited state $|\varphi^{e_e}_A\rangle$ of $H^e_A$.
The experimental fidelities are $\mathcal F^\mathrm{exp}=99.2\%$ and $96.4\%$ while their theoretical predictions are both $\mathcal F^\mathrm{theo}=99.8\%$.
(iv) The above iteration stages lead to two product states,
$\{|\psi_{\gamma=1,2}\rangle=|\varphi^{g_g}_A\rangle\otimes|\varphi^g_B\rangle, |\varphi^{e_e}_A\rangle\otimes|\varphi^e_B\rangle\}$.
Following a symmetry argument, the other two product states $\{|\psi_{\gamma=3,4}\rangle\}$
are obtained for the cluster partition of $A^\pr=\{\mathrm{spins}~2, 3\}$ and $B^\pr=\{\mathrm{spin}~1\}$.
The subsequent Schmidt orthogonalization gives rise to four orthogonal
basis states $\{\psi^\mathrm{S}_{\gamma=1,\cdots, 4}\rangle\}$ and a 4-D effective Hamiltonian $H_\mathrm{eff}$.
As shown in Fig.~\ref{fig_02}(e), the experimental determination of $|\Psi^\mathrm{exp}_g\rangle$ is converged over $70\sim100$ VQE steps,
with a high fidelity $\mathcal F^\mathrm{exp}_g=95.4\%$  as compared to the theoretical prediction $\mathcal F^\mathrm{theo}_g=99.3\%$.

\begin{figure}[tp]
\centering
 \includegraphics[width=0.7\columnwidth]{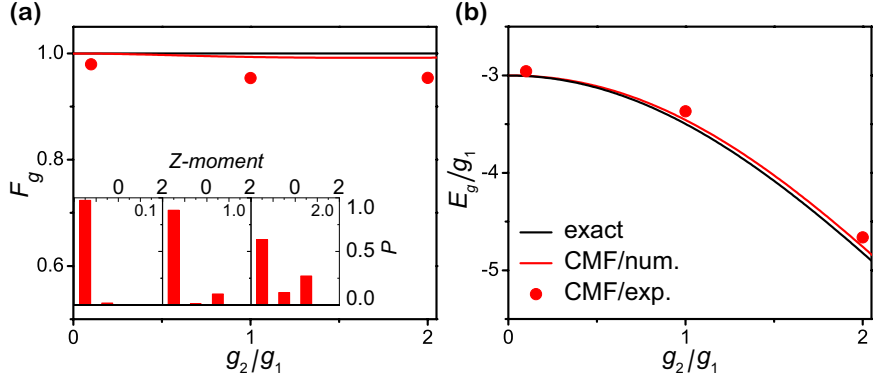}
\caption{The CMF determination of (a) the fidelity of the ground state $|\Psi_g\rangle$ and (b) the corresponding eigenenergy $E_g$
for the three-spin Hamiltonian in Eq.~(\ref{eq_06}) with a fixed $g_3/g_1=0.1$ and a varying $g_2/g_1$.
The red lines and circles denote the numerical and experimental results via the CMF method while the black lines denote the exact values.
In the inset of (a), the distributions of the $Z$-moment are shown for the experimentally determined $|\Psi^\mathrm{exp}_g\rangle$
with $g_2/g_1=0.1$, 1.0 and 2.0.
}
\label{fig_03}
\end{figure}

In our numerical calculation of $N$-spin systems,
the total 8 product states are considered in the construction of the compressed Hilbert space.
In the first stage of our experiment, we only consider the ground state of $H_A$
so that the total 4 product states arisen from its first excited state are excluded.
The inset in Fig.~\ref{fig_02}(e) shows a schematic diagram of the effective Hamiltonian,
from which we find that $|\Psi_g\rangle$  can be obtained from the 3-D or  2-D spaces with
$\mathcal F^\mathrm{theo}_g=99.3\%$ and 99.0\%. Thus, a continued compression over the product states
is allowed to further decrease the cost of a CMF eigensolver.

In Fig.~\ref{fig_03}, we present the experimental results of $|\Psi^\mathrm{exp}_g\rangle$
and $E^\mathrm{exp}_g$ for a fixed $g_3/g_1=0.1$
and a varying $g_2/g_1$ ($=0.1$, 1.0 and 2.0) based on the CMF algorithm. As compared to
the exact ground state, the theoretical predictions of the state fidelity is excellent
($\mathcal F^\mathrm{theo}_g>99\%$) while the experimental results are consistently high,
$\mathcal F^\mathrm{exp}_g=97.9\%$, $95.4\%$ and $95.4\%$ [see Fig.~\ref{fig_03}(a)].
The same behavior is found for the accuracy of $E^\mathrm{exp}_g$ [see Fig.~\ref{fig_03}(b)].
In a simplified scenario of $g_3=0$,  this three-spin system  prefers ferromagnetism along the
$Z$-direction for $g_2/g_1\rightarrow 0$ while anti-ferromagnetism along the
$X$-direction in the opposite limit ($g_2/g_1\rightarrow \infty$). The ground state thus experiences a transition
from $|\Psi_g(g_2/g_1\rightarrow 0)\rangle=|111\rangle$ to
$|\Psi_g(g_2/g_1\rightarrow \infty)\rangle\!\propto\!\prod \limits_{i=1}^3(|0\rangle-(-1)^i|1\rangle)_i$.
Here we introduce the total spin moment $M_j=\sum_{i=1}^N m_{i; j}$, where $m_{i; j}$ is the magnetic moment
of each $i$-th spin along the $j(=X, Y, Z)$-direction.
In the parameter range in our experiment, the entanglement of $|\Psi_g\rangle$ increases with $g_2/g_1$,
indicated by a broadening distribution of the $Z$-moment $M_Z$ in the inset of Fig.~\ref{fig_03}(a).

\begin{figure}[tp]
\centering
 \includegraphics[width=0.7\columnwidth]{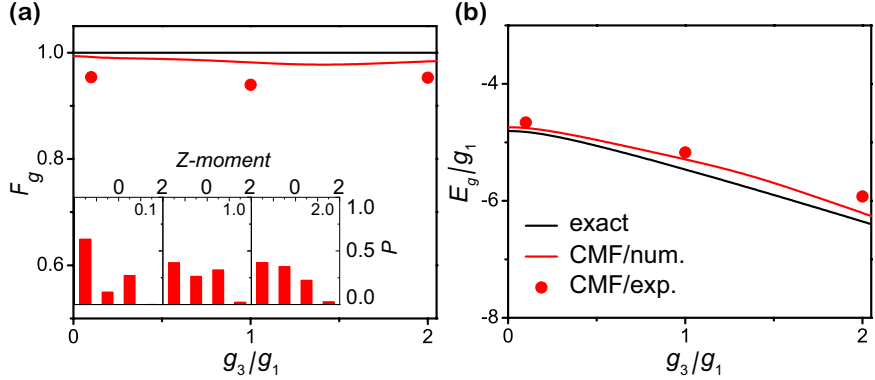}
\caption{The CMF determination of (a) the fidelity of the ground state $|\Psi_g\rangle$ and (b) the corresponding eigenenergy $E_g$
for the three-spin Hamiltonian in Eq.~(\ref{eq_06}) with a fixed $g_2/g_1=2.0$ and a varying $g_3/g_1$.
The red lines and circles denote the numerical and experimental results via the CMF method while the black lines denote the exact values.
In the inset of (a), the distributions of the $Z$-moment are shown for the experimentally determined $|\Psi^\mathrm{exp}_g\rangle$
with $g_3/g_1=0.1$, 1.0 and 2.0.}
\label{fig_04}
\end{figure}

In our second set of experiments, we fix  $g_2/g_1=2.0$ and consider three values of $g_3/g_1=0.1$, 1.0 and 2.0.
As shown in the inset of Fig.~\ref{fig_04}(a), the increase of $g_3/g_1$ also leads to an extensive distribution of the $Z$-moment.
The same CMF algorithm as in Fig.~\ref{fig_02} reliably determines the ground states. The experimental results of
the state fidelities are presented in Fig.~\ref{fig_04}(a), satisfying $\mathcal F^\mathrm{exp}_g=95.4\%$, $94.0\%$ and $95.3\%$
for the three input parameters. The accuracy of the experimentally extracted eigenenergy
$E^\mathrm{exp}_g$ follows the same trend [see Fig.~\ref{fig_04}(b)].

{\it Summary.} --- In this paper, we apply a multi-layer CMF method to design a new quantum eigensolver so that the eigenstates
of a large-scale quantum system can be determined by a series of quantum computations over its clusters.
For a pre-selected cluster, certain eigenstates of its reduced Hamiltonian are extracted via a quantum algorithm
after a partial average over an eigenstate of the environment cluster. The products
of eigenstates from different clusters are used to construct a compressed Hilbert space,
in which the effective Hamiltonian is digonalized to determine certain eigenstates of the whole Hamiltonian.
This CMF method is numerically verified in the $N$-spin chains with two-spin interactions. For the condition of
$3\le N\le 8$, the CMF calculations in the $2^{N^\pr}(N^\pr\le4)$-D spaces provide an excellent prediction on the ground state
of the $2^N$-D Hamiltonian with  $\mathcal F^\mathrm{theo}_g>99.4\%$.
This CMF method is further experimentally studied in the 3-spin chain with both two- and three-spin interactions.
Under various parameter combinations, the experimental determination of the ground state  via the CMF method
is consistently high, satisfying $\mathcal F^\mathrm{exp}_g\gtrsim 95\%$. The studies of the ground states in this paper
can be straightforwardly extended to the excited states.
With a quick size increment of quantum devices~\cite{GoogleNat19,GongSCI21}, the CMF method shows its promise to sufficiently large-scale
Hilbert spaces.

\begin{acknowledgments}
The work reported here was supported by the National Key Research and Development
Program of China (Grant No. 2019YFA0308602, No. 2016YFA0301700), the National Natural
Science Foundation of China (Grants No. 12074336, No. 11934010, No. 11775129), the
Fundamental Research Funds for the Central Universities in China (2020XZZX002-01), and the Anhui Initiative
in Quantum Information Technologies (Grant No. AHY080000). Y.Y. acknowledge the
funding support from Tencent Corporation. This work was partially conducted at the University
of Science and Technology of China Center for Micro- and Nanoscale Research
and Fabrication.

\end{acknowledgments}

\end{document}